\documentclass[Afour,times,sagev]{sagej}

\usepackage{moreverb,url}
\usepackage{bm}
\usepackage[colorlinks,bookmarksopen,bookmarksnumbered,citecolor=red,urlcolor=red]{hyperref}
\usepackage{stfloats}
\usepackage{algorithm}
\usepackage{algpseudocode}
\usepackage{setspace}
\usepackage{amsmath}
\usepackage{multirow}
\newcommand\BibTeX{{\rmfamily B\kern-.05em \textsc{i\kern-.025em b}\kern-.08em
T\kern-.1667em\lower.7ex\hbox{E}\kern-.125emX}}

\def\volumeyear{2018}

\begin{document}
\runninghead{Xuejie Mai, Zhiyong Yuan, Qianqian Tong et al.}

\title{Periodic-corrected data driven coupling of blood flow and vessel wall for virtual surgery}

\author{Xuejie Mai, Zhiyong Yuan, Qianqian Tong, Tianchen Yuan and Jianhui Zhao}

\affiliation{School of Computer, Wuhan University, PR China}

\corrauth{Zhiyong Yuan, School of Computer Science, Wuhan University, Wuhan, Hubei 430072, P.R. China.}

\email{zhiyongyuan@whu.edu.cn}

\begin{abstract}
Fast and realistic coupling of blood flow and vessel wall is of great importance to virtual surgery. In this paper, we propose a novel data-driven coupling method that formulates physics-based blood flow simulation as a regression problem, using an improved periodic-corrected neural network (PcNet), estimating the acceleration of every particle at each frame to obtain fast, stable and realistic simulation. We design a particle state feature vector based on smoothed particle hydrodynamics (SPH), modeling the mixed contribution of neighboring proxy particles on the blood vessel wall and neighboring blood particles, giving the extrapolation ability to deal with more complex couplings. We present a semi-supervised training strategy to improve the traditional BP neural network, which corrects the error periodically to ensure long term stability. Experimental results demonstrate that our method is able to implement stable and vivid coupling of blood flow and vessel wall while greatly improving computational efficiency.
\end{abstract}


\maketitle
\bibdata{2}
\section{Introduction}
Virtual surgery is an important application of virtual reality in the medical field. Through the virtual surgery system, doctors can carry out clinical diagnosis, surgical training and surgical planning to assist in the development of rational surgical programs to improve the success rate of surgery.\cite{1} Due to the interaction between blood flow and the vessel wall, hemodynamic factors profoundly affect the quality of many surgeries.\cite{2,3} On the one hand, the hemodynamics of blood vessels, especially arterial blood vessels, are closely related to vascular diseases such as intimal thickening and atherosclerotic plaque formation.\cite{4} On the other hand, intravascular hemodynamics is strongly sensitive to the geometry of blood vessels. Small deformations and shearing of blood vessel walls during surgery lead to significant differences in hemodynamics. Therefore, the fast and realistic blood vessel wall coupling is of great significance to improve the realism and immersion of virtual surgery. At present, the research of virtual surgery system realizes the real-time interactive visual haptic model between human soft tissues and virtual surgical instruments.\cite{5} However, the real-time simulation of realistic blood flow is challenging, especially large-scale simulation based on the physical size.

This paper aims to introduce a data-driven approach into SPH-based blood flow simulation to achieve fast and realistic coupling of blood flow and vessel wall. A first thought is to input the position and velocity of each particle in the current frame into the neural network to output the position and velocity of each particle in the next frame. However, the relationship of position and velocity of each particle between frames is unstable, which is easily affected by factors such as time step and external force. In contrast, the relationship between the input and output data in the acceleration calculation step is relatively stable. Thus, we design a feature vector based on neighbor information of particles in the current frame to predict the acceleration of particles in the next frame to speed up the acceleration calculation step and the entire coupling simulation process.

In the following sections, we first describe the traditional SPH-based method of blood flow simulation. Then, on the basis of the SPH method and the hemodynamic model, we construct a feature vector of the blood flow state to mimic the mixed contribution of neighborhoods composed of proxy particles on the vessel wall and blood flow particles, giving the extrapolation ability to deal with more complex couplings. Finally, we propose a semi-supervised neural network training strategy that periodically corrects errors of the traditional BP neural network to ensure long-term stability. As shown in Figure 1, the experimental results show that the proposed data-driven method based on a periodic-corrected neural network (PcNet) realizes the steady and realistic coupling of blood flow and vessel wall in virtual surgery and greatly improves the simulation efficiency.
\begin{figure*}
\centering
\includegraphics[width=5in]{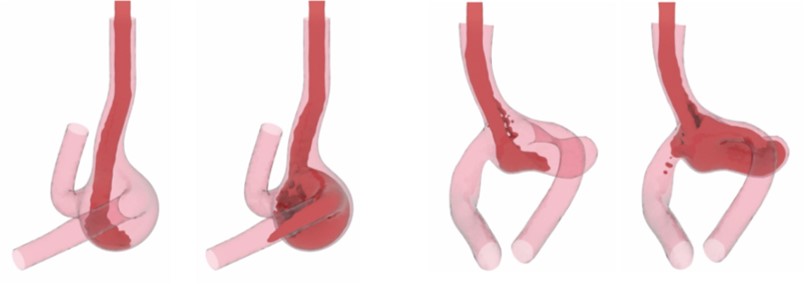}
\caption{The simulation results using our data-driven coupling method for blood flow and vessel wall.
The proposed method achieves almost the same simulation visual effects as the physics-based method, and even better on the details of the blood splashes, while the computational efficiency is improved by about 5 times.}
\end{figure*}
\section{Related Work}
Fluid simulation is a long standing problem in computer graphics and virtual reality. In recent years, many researchers have proposed a number of physics-based methods to simulate the behavior of fluids and the interactions between fluids and rigid or deformable solids. The major drawback of grid-based approaches is their computational complexity, which limits their use in interactive virtual surgery.

Particle-based methods are widely used in fluid simulation with borderline adaptability and good computational performance.\cite{6,7,8,9,10} However, the research on the direction of blood flow simulation is scarce. M\"uller et al. simplified the blood flow to Newtonian fluid for coupling simulation.\cite{11} Qin et al. implemented the method of M\"uller et al. using PPU parallel acceleration.\cite{12} Jing et al. coupled SPH and particle spring models to simulate blood flow.\cite{13} The above method lacks visual realism due to neglecting the non-Newtonian fluid characteristics of blood flow, and at the same time the simulation is small in scale and the real-time performance is not sufficient. Lai et al. implemented GPU-assisted blood flow simulation using the Compute Unified Device Architecture (CUDA), achieving 50 fps simulation at 9,000 particles, but did not consider the interaction of blood flow with blood vessels.\cite{14} Guo et al. proposed a GPU-accelerated mixed particle-based coupling method for blood flow and vessel wall that real-time simulated at 100,000 particle size.\cite{15} Although these studies have made some improvements in computational efficiency, a more efficient method is urgently needed due to the multi-phase coupling involved in the virtual surgery scene with flexible organs and surgical instruments.

In recent years, data-driven methods are increasingly proposed in the field of fluid simulation. Raveendran et al. generated a large number of fluid simulations by interpolating the existing fluid simulation results.\cite{16} Yang et al. used neural network prediction instead of the diffusion projection in the gridding method to achieve a speed increase of about 10 times.\cite{17} Jeong et al. used regression forests to simulate fluid particle state changes, which can achieve 10-1000 times faster operation on the GPU.\cite{18} At present, there are few studies that combine data-driven methods with blood flow simulation and even virtual surgery.
\section{1 Basic method}
\subsection{1.1 SPH method and boundary sampling}
The key idea of the SPH method is that the physical quantity $A({{\mathbf{r}}_{i}})$  at the position ${\mathbf{r}}_{i}$ in the fluid is obtained by adding the corresponding physical quantities of the neighboring particles under the action of the smooth kernel function:
\begin{equation}
A({{\mathbf{r}}_{i}})=\sum\limits_{j}{{{A}_{j}}\frac{{{m}_{j}}}{{{\rho }_{j}}}W(}{{\mathbf{r}}_{i}}-{{\mathbf{r}}_{j}},h)
\end{equation}where $j$ donates the particle within a distance of $h$ from particle $i$ (namely neighboring particle), $A$ is the physical quantity of particle, $m$ is the mass, $\rho$ is the density, and $W$ is a spherically symmetric smooth kernel function.
When dealing with the coupling of blood flow vessel walls, the proxy particles on the vessel wall surface are sampled and included in the neighborhood search range for the density calculation of the border blood flow particle $i$:
\begin{equation}
\rho_{i} = m_{i}\sum\limits_j {W(} {\mathbf{r}_{ij}},h) + \sum\limits_k {{\varpi _{{s_k}}}W(} {\mathbf{r}_{ik}},h)
\end{equation}
where $j$ and $k$ are the blood flow particles and the proxy particles in the neighborhood respectively. The contribution of each proxy particle to the blood flow particle density is expressed as ${\varpi _{{s_i}}=\rho_{0} V_{{s_i}}}$, where the volume of the proxy particle $V_{{s_i}}$ can be defined as:
\begin{equation}
V_{{s_i}} = \frac{1}{{\sum\limits_k {W(} \mathbf{r} - {\mathbf{r}_k},h)}}
\end{equation}
\subsection{1.2 Hemodynamic equation}
Blood flow is essentially a non-Newtonian fluid, thus we describe the blood flow using the Navier-Stokes equation that includes non-Newtonian terms as follows:\cite{15}
\begin{equation}
\frac{{\partial \rho }}{{\partial t}}\; + \;\nabla  \cdot (\rho \mathbf{v}) = 0
\end{equation}
\begin{equation}
\mathbf{f} = \rho \mathbf{a} =  - \nabla p + \nabla  \cdot {\bm{\tau}}  + \rho \mathbf{g}
\end{equation}
Equation (4) describes the mass conservation of blood flow (SPH naturally satisfies this equation) and Equation (5) illustrates the momentum conservation. In the above equations, $\mathbf{a}$ is the acceleration of the blood flow particle, and the physical quantities to be solved are density $\rho$, pressure term $- \nabla p$, viscous force term $\nabla\cdot{\bm{\tau}}$ and external force term $\rho \mathbf{g}$. We describe $\bm{\tau}$ through strain rate tensor $\dot{\bm{\gamma}}$:
\begin{equation}
\bm{\tau} \;{\rm{ = }}\;\upsilon \dot{\bm{\gamma}}
\end{equation}
\begin{equation}
\dot{\bm{\gamma}} \;{\rm{ = }}\;(\nabla \mathbf{v}\; + \;{(\nabla \mathbf{v})^T})/2
\end{equation}
where $\mathbf{v}$ is the velocity, $\upsilon$ is the viscosity, which donates a function of the second invariant ${D}_{\Pi}$ at the shear rate as an independent variable and can be described by the constitutive equation of the Casson model:
\begin{equation}
\upsilon ({{D}_{\Pi}})=\frac{[\sqrt{\eta }\sqrt{\sqrt{2{{D}_{\Pi}}}}+\sqrt{{{\tau}_{y}}}1-{{e}^{-n\sqrt{2{{D}_{\Pi}}}}}){{]}^{2}}}{\sqrt{2{{D}_{\Pi}}}}
\end{equation}
\begin{equation}
\eta\;{\rm{ = }}\;\rho \mu
\end{equation}
In Equation (8), $\mathbf{\tau}_{y}$ is the shear yield stress, $\eta$ donates the Casson viscosity, $1-{{e}^{-n |\dot{\bm{\gamma}} |}}$ is introduced by avoiding $\upsilon$ becoming a singular value as $\bm{\gamma}$ approaches zero, $n$ is a finite constant value(here set $n=7$).
\subsection{1.3 Numerical solution}
We first calculate the velocity gradient $\nabla {\mathbf{v}_i}$ according to the SPH formula:
\begin{equation}
\nabla {\mathbf{v}_i} = \sum\limits_j {\frac{{{m_j}}}{{{\rho _j}}}\nabla W(} {\mathbf{r}_{ij}},h){\mathbf{v}_{ji}}^T
\end{equation}
Calculate $\bm{\tau}_i$ according to Equations (6) (7) (8), and then use the SPH method to get the sum of the viscous force item ${\mathbf{f}_i}^v$ and pressure item ${\mathbf{f}_i}^p$, which is the fluid force ${\mathbf{f}_{i \leftarrow j}}$:
\begin{equation}
{\mathbf{f}_{i \leftarrow j}} = {\mathbf{f}_i}^v + {\mathbf{f}_i}^p
\end{equation}
\begin{equation}
{\mathbf{f}_i}^v = \sum\limits_j {\frac{{{m_j}}}{{{\rho _i}{\rho _j}}}({\bm{\tau} _i} + {\bm{\tau} _j})\nabla W(} {\mathbf{r}_{ij}},h)
\end{equation}
\begin{equation}
{\mathbf{f}_i}^p =  - \sum\limits_j {{m_j}(\frac{{{p_i}}}{{{\rho _i}^2}} + \frac{{{p_j}}}{{{\rho _j}^2}})\nabla W(} {\mathbf{r}_{ij}},h)
\end{equation}
where $p_i$ is calculated by Tait equation:
\begin{equation}
{p_i} = \frac{{{\rho _0}{c_s}^2}}{\gamma }({(\frac{{{\rho _i}}}{{{\rho _0}}})^\gamma } - 1)
\end{equation}
where $c_s$ is the speed of the sound in the fluid. External forces, including gravity, act directly on the blood flow particles, eliminating the need for the SPH method.

In the process of coupling, the coupling force of the proxy particle to the blood flow particle is as follows:
\begin{equation}
{\mathbf{f}_{i \leftarrow {\rm{k}}}} =  - {m_i}{\varpi _{{s_k}}}((\frac{{{p_i}}}{{{\rho _i}^2}} + \frac{{{p_k}}}{{{\rho _k}^2}} + {\widetilde \Pi _{ik}})\nabla {W_{ik}}
\end{equation}
where $k$ represents the proxy particle, similar to the Equation (2) with ${\varpi _{{s_i}}=\rho_{0} V_{{s_i}}}$ to modify the contribution of the proxy particle. Viscosity term ${\widetilde \Pi _{ik}}$ is defined as:\cite{19}
\begin{equation}
{\widetilde \Pi _{ik}}{\rm{ = }} - \frac{{16{\mu _i}{\mu _k}}}{{{\rho _i}{\rho _k}({\mu _i} + {\mu _k})}}(\frac{{{\mathbf{v}_{ik}} \cdot {\mathbf{r}_{ik}}}}{{{\mathbf{r}_{ik}} + \varepsilon {h^2}}})
\end{equation}
where $\mathbf{v}_{ik}$ and $\mathbf{r}_{ik}$ donate the relative velocity and the relative position between particles respectively, including viscosity coefficient ${{\mu }_{i}}=\upsilon (D)h{{c}_{s}}/{{p}_{i}}$, with $\varepsilon =0.01{{h}^{2}}$ to prevent the generation of singular value $|{{r}_{ij}}|\text{= }0$. Depending on the ideal state of the incompressible fluid, $p_k$, $\rho_k$ and $\mu_k$ of the proxy particle $k$ take the same value as the blood flow particle $i$.

The resultant fluid force, external force, and coupling force are summed to obtain the acceleration ${\mathbf{a}_i}^{n + 1}$ of the next frame. Finally, the velocity and position of particle are updated according to Equations (17) and (18):
\begin{equation}
\mathbf{v}_i^{n + 1} = \mathbf{v}_i^n + {\mathbf{a}_i}^{n + 1}\Delta t
\end{equation}
\begin{equation}
\mathbf{x}_i^{n + 1} = \mathbf{x}_i^n + \frac{{\mathbf{v}_i^n + \mathbf{v}_i^{n + 1}}}{2}\Delta t
\end{equation}

In summary, physically SPH-based coupling method for blood flow and vessel wall as shown in Algorithm 1.

For each blood flow particle $i$ in the current frame $n$, first search for its neighboring particles. Then, calculate the force $\mathbf{f}_i$ according to the velocity ${\mathbf{v}_j}^n$ and position ${\mathbf{r}_j}^n$ of the neighboring blood flow particle $j$ and the velocity ${\mathbf{v}_k}^n$ and position ${\mathbf{r}_k}^n$ of the neighboring proxy particle $k$ to obtain the acceleration ${\mathbf{a}_i}^{n+1}$. Finally utilize the time integration equations to solve the velocity ${\mathbf{v}_i}^{n+1}$, position ${\mathbf{r}_i}^{n+1}$ in the next frame and update the blood flow state. In the second step above, the acceleration calculation based on the mixed neighborhood information of the particles is very complicated, and the overall relationship between the input and output data in the step is relatively stable, without being affected by factors such as time steps and external forces.
\begin{algorithm}[h]
  \caption{Physically SPH-based coupling of blood flow and vessel wall}
  \begin{algorithmic}[1]
   \While{$animating$}
        \For{each blood particle $i$}
        \State find neighboring particles $j$ and $k$;
        \State compute fluid forces $\mathbf{f}_{i \leftarrow j}$, (11);
        \State compute coupling forces $\mathbf{f}_{i \leftarrow k}$, (15);
        \State compute acceleration ${\mathbf{a}_i}^{n+1}$, (5);
        \State update velocity ${\mathbf{v}_i}^{n+1}$, position ${\mathbf{r}_i}^{n+1}$, (17)(18);
        \EndFor
   \EndWhile
  \end{algorithmic}
\end{algorithm}

Therefore, we construct a particle state feature vector based on the neighborhood information of the current frame to predict the acceleration of the next frame, accelerating this step and the entire coupling simulation. It also offers the extrapolation capability of simulating more complex couplings.
\section{2 Our method}
As shown in Algorithm 2, we propose a data-driven coupling method for blood flow and vessel wall. After neighborhood search, the feature vector is extracted from the blood flow state data of the current frame and is input into the trained PcNet to predict and obtain the acceleration to update the blood flow state of the next frame. In the following, the particle state feature vector based on the mixed neighborhood and the semi-supervised periodic-corrected training strategy for PcNet are described in detail.
\begin{algorithm}
  \caption{Data-driven coupling of blood flow and vessel wall}
  \begin{algorithmic}[1]
   \While{$animating$}
        \For{each blood particle $i$}
        \State find neighboring particles $j$, $k$;
        \State extract feature vector $F(a_{i}^{n},{{r}^{n}},{{v}^{n}})$;
        \State predict acceleration ${\mathbf{a}_i}^{n+1}$ using PcNet;
        \State update velocity ${\mathbf{v}_i}^{n+1}$, position ${\mathbf{r}_i}^{n+1}$, (17)(18);
        \EndFor
   \EndWhile
  \end{algorithmic}
\end{algorithm}
\subsection{2.1 Particle state feature vector}
Focusing on the acceleration calculation of the physically SPH-based coupling method, especially the calculation formulas of physical quantities such as density $\rho _i$, fluid force $\mathbf{f}_{i \leftarrow j}$ and coupling force $\mathbf{f}_{i \leftarrow k}$, we can easily find out that it is the information of the position and velocity of each blood flow particle and its neighboring particles that affect the particle state in next frame, which is the original features of each blood flow particle.

We note that different particles have different numbers of neighboring particles and the number of neighboring particles of a certain particle is also changing in the time series. To deal with the different dimensions of the original particle features, we use the basic statistical methods to measure the distribution of data (mean, variance, skewness and kurtosis) to extract the features of neighboring particle position and velocity distribution of blood flow particles. Noting the difference between computing fluid forces and coupling forces, we measure features of neighboring blood flow particles and neighboring proxy particles respectively. In addtion, we add the number of neighboring blood flow particles $N$ and the number of neighboring proxy particles $M$ to the feature vector.
\begin{figure}[H]
\centering
\includegraphics[width=3in]{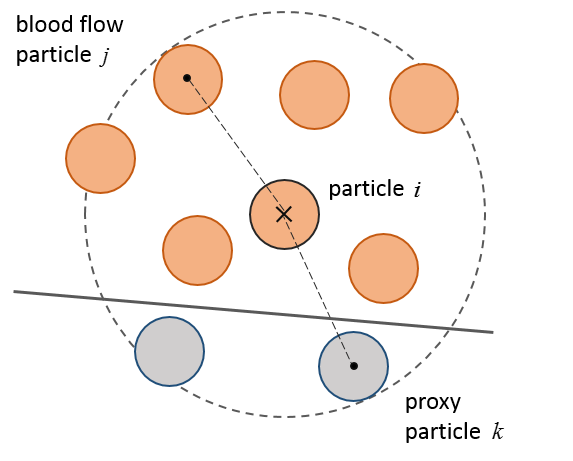}
\caption{The neighborhood contains blood flow particles and proxy particles.}
\end{figure}
Assuming that the velocity of the blood flow particle in the current frame is ${{v}_{i}}=({{v}_{xi}},{{v}_{yi}},{{v}_{zi}})$, the position is ${{r}_{i}}=({{r}_{xi}},{{r}_{yi}},{{r}_{zi}})$, $j$ is the neighboring blood flow particle, and $k$ is the neighboring proxy particle as shown in Figure 2. In the following, we take the neighboring blood particle position and velocity distribution feature extraction as an example.

\subsubsection{1. Central tendency feature  }
We use the arithmetic mean of relative values to characterize the central tendency of the velocity and position of the neighboring particles. The average of the $x$-component differences in the positions of the neighboring blood flow particles is obtained as:
\begin{equation}
dis{x_{avg}}^b = \frac{1}{N}\sum\limits_j {({r_x}_j - {r_{xi}}} )
\end{equation}

The average of the $x$-component differences in the velocities of the neighboring blood flow particles can be calculated as:
\begin{equation}
rv{x_{avg}}^b = \frac{1}{N}\sum\limits_j {({v_x}_j - {v_{xi}}} )
\end{equation}

Then we calculate $dis{y_{avg}}^b$(average of the $y$-component in the relative positions), $dis{z_{avg}}^b$(average of the $z$-component in the relative positions), $rv{y_{avg}}^b$(average of the $y$-component in the relative velocities) and $rv{z_{avg}}^b$(average of the $z$-component of the relative velocities).
\subsubsection{2. Discrete degree feature  }
We use the variance of relative values to characterize the discreteness of the velocity and position of the neighboring particles. The variance of the relative positions of the neighboring blood flow particles is obtained as:
\begin{equation}
{D_{dis}}^b = \frac{1}{N}\sum\limits_j {(|{\mathbf{r}_{ij}}| - \overline{|\mathbf{r}_{ij}|} } {)^2}
\end{equation}
where $\overline{|\mathbf{r}_{ij}|}$ denotes the average distance between the neighboring blood flow particle $j$ and the particle $i$.
The variance of the relative velocities of the neighboring blood flow particles can be calculated as:
\begin{equation}
{D_{rv}}^b = \frac{1}{N}\sum\limits_j {(|{\mathbf{v}_{ij}}| - \overline{|\mathbf{v}_{ij}|} } {)^2}
\end{equation}
where $\overline{|\mathbf{v}_{ij}|}$ denotes the average value of the relative velocities of the neighboring blood flow particles.
\subsubsection{3. Distribution shape feature  }
We use the skewness coefficient and kurtosis coefficient of relative values to characterize the distribution shape of the velocity and position of the neighboring particles. Note that the calculation of skewness coefficient and kurtosis coefficient depends on frequency distribution.

Taking the relative position of neighboring particles as an example, it should be discretized first. Suppose the smooth core radius of the particle is $h$, thus the range of the distance of the blood flow particle in the neighborhood is $[0,h]$. We divide this range into $n$ aliquots (here set $n=6$) and the distance of the particle in the $i$-th partition is defined as ${l_i}=i*h/n$.Then we count the frequency $F_{l_i}$ of particles in each partition and apply the moment method to measure skewness coefficient and kurtosis coefficient. In general, taking the center point $a$ of data $X$, the $n$-th moment of $X$ with respect to $a$ is defined as $\sum{{{(X-a)}^{k}}}/N$.

In statistics, the skewness coefficient is used to measure the direction and degree of inclination of data frequency distribution, which is generally calculated by the third-order center moment. The kurtosis coefficient is used to measure curve sharp or flat top level of data frequency distribution, which is usually calculated by the fourth-order center moment. We can formulate the skewness coefficient of the relative position distribution of the neighboring blood flow particles as:
\begin{equation}
{\alpha _{dis}}^b = \frac{{\sum\limits_j^{} {{{({l_j} - \overline l )}^3}{F_j}} }}{{{\sigma _l}^3\sum\limits_j^{} {{F_j}} }}{\rm{ = }}\frac{{\sum\limits_j^{} {{{({l_j} - \overline l )}^3}{F_j}} }}{{{\sigma _l}^3N}}
\end{equation}
where $\sigma _l$ stands for the standard deviation. The kurtosis coefficient of the relative position distribution of neighboring blood flow particles is obtained as:
\begin{equation}
{\beta _{dis}}^b = \frac{{\sum\limits_j^{} {{{({l_j} - \overline l )}^4}{F_j}} }}{{{\sigma _l}^4\sum\limits_{i = 1}^{} {{F_j}} }} = \frac{{\sum\limits_j^{} {{{({l_j} - \overline l )}^4}{F_j}} }}{{{\sigma _l}^4N}}
\end{equation}

Then we calculate the skewness coefficient ${\alpha _{rv}}^b$ and the kurtosis coefficient ${\beta _{rv}}^b$ of the relative velocity distribution of the neighboring blood flow particles.

In the same way, the position and velocity distribution features of proxy particles on the blood vessel wall can be extracted. In this paper, the blood vessel model is fixed, which means the velocity of proxy particles is zero and its velocity distribution features does not need to be extracted. The extracted features are as follows:	
\begin{equation}
\{\,dis{{x}_{avg}}^{p},dis{{y}_{avg}}^{p},dis{{z}_{avg}}^{p},{{D}_{dis}}^{p}{{\alpha }_{dis}}^{p},{{\beta }_{dis}}^{p}\,\}	
\end{equation}

Considering the dynamic frame correlation of blood flow simulation, we add the acceleration in the current frame ${a_i^n}=\{a_x, a_y, a_z\}$ to the feature vector. In summary, the particle state feature vector based on the mixed neighborhood of proxy particles on the blood vessel wall and blood particles is constructed as a 23-dimensional vector:
\begin{align}
\begin{split}
  & \mathbf{F} = \,\!\!\{\,{{a}_{x}},\text{ }{{a}_{y}},\text{ }{{a}_{z}}, \\
 & \qquad N,dis{{x}_{avg}}^{b},dis{{y}_{avg}}^{b},dis{{z}_{avg}}^{b}, \\
 & \qquad rv{{x}_{avg}}^{b},\ rv{{y}_{avg}}^{b}rv{{z}_{avg}}^{b}, \\
 & \qquad {{D}_{dis}}^{b},{{D}_{rv}}^{b},\ {{\alpha }_{dis}}^{b},{{\alpha }_{rv}}^{b},{{\beta }_{dis}}^{b},{{\beta }_{rv}}^{b} \\
 & \qquad M,dis{{x}_{avg}}^{p},dis{{y}_{avg}}^{p},dis{{z}_{avg}}^{p}, \\
 & \qquad {{D}_{dis}}^{p}{{\alpha }_{dis}}^{p},{{\beta }_{dis}}^{p}\,\}\!\! \\
 \end{split}
\end{align}
\subsection{2.2 Traditional neural network}
The classic BP (Back Propagation) neural network is a computational network structure composed of many neurons. Figure 3 shows a classic three-layer BP neural network, including the input layer, hidden layer and output layer. A neuron is a basic data processing unit in the network. The working mechanism contains two phases: forward propagation (right black arrow) and backward propagation (left yellow arrow). In the forward propagation phase, the values of the upper nodes are weighted according to the weights of the corresponding directed arcs. After the bias is added, the output value is obtained through the activation function to be transmitted to the lower nodes. In the reverse propagation phase, according to the gradient descent method constantly update weights and bias to minimize errors between output and target values.\cite{20}

The basic neural network in this paper takes the 23-dimensional particle state feature vector $\mathbf{F}$ of the current frame as the input layer and the 3-dimensional particle acceleration $\mathbf{T}=\{a_i^{n+1}\}$ in the next frame as the output layer. The input layer contains 24 neurons (including the biased neuron) and the output layer contains 3 neurons. The rest of the network structure settings (the number of hidden layers and the number of neurons per layer) need to be experimentally adjusted, which is not discussed here.
\begin{figure}[H]
\centering
\includegraphics[width=2.75in]{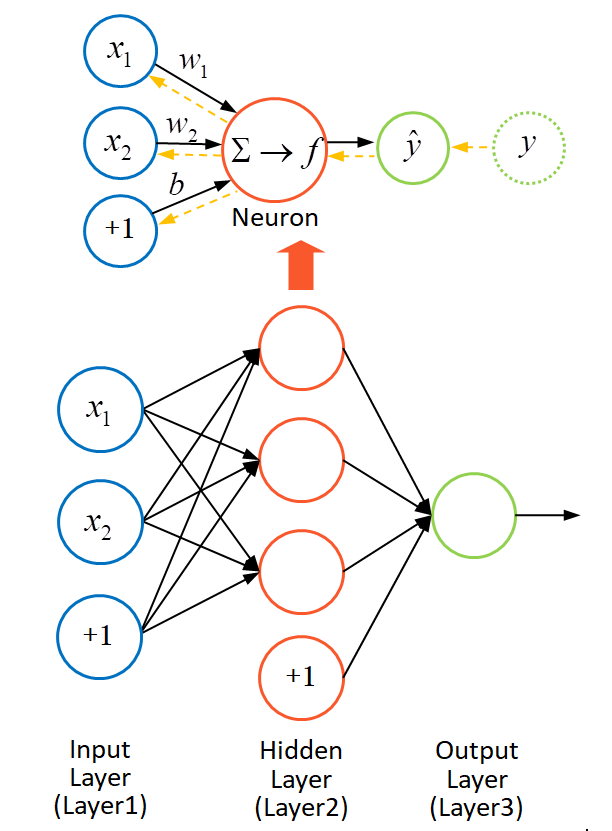}
\caption{An example of a three-layer BP neural network}
\end{figure}
\subsection{2.3 Periodic-corrected training strategy}
The entire blood flow simulation process consists of a continuous sequence of frames; the frame sequence has a dynamic continuous correlation. We find that training with classic BP neural networks can achieve a sufficiently low error in a single time step; however, the error will continue to accumulate over successive frame-by-frame predictions leading to instabilities. This is precisely because the training strategy of classic BP neural network only focuses on the prediction errors of adjacent frames, making it impossible to give consideration to the continuous physical meaning of blood flow simulation. Therefore, the traditional neural network¡¯s architecture has to be improved in order to enhance its applicability based on time-series blood flow simulation.
\begin{algorithm}[H]
  \caption{Periodic-corrected training strategy}
  \begin{algorithmic}[1]
    \Require
      feature vectors $\bm{F}$, target vectors $\bm{T}$.
    \Ensure
      trained periodic-corrected network $PcNet$.
    \For{each training epoch}
    \For{each training sample $i$}
      \State $PcNet$.trian$(\bm{F}_i,\bm{T}_i)$;
      \If {$i$ mod $a$ == $0$}
        \State $PcNet$.predict:$(\bm{F}_1,\bm{T}_1) \rightarrow (\bm{F}_i,\bm{T}_i)$;
        \State $PcNet$.train$(\hat{\bm{F}_1},\bm{T}_1)$;
      \EndIf
    \EndFor
    \EndFor
  \end{algorithmic}
\end{algorithm}
As shown in Algorithm 3, we propose a semi-supervised neural network training strategy that periodically corrects errors to train our PcNet model. Suppose a frame $(\mathbf{F},\mathbf{T})$ is a training sample. The training samples are extracted from the physical flow simulation to form a continuous sequence (physical sequence), and the sequence is continued for iterative training. Given a period length $a$ (here set $a=5$), the iterative algorithm for improved periodic-corrected training is as follows:
\begin{enumerate}
\item[(1)] Train $a-1$ training samples, adjust and update the weights and biases of the network to get $PcNet_1$.
\item[(2)] Utilize $PcNet_1$ to perform continuous frame-by-frame prediction from the first frame $(\mathbf{F}_1,\mathbf{T}_1)$ to form a prediction sequence and obtain the predicted value of the $a$-th frame $(\hat{\mathbf{F}}_i,\hat{\mathbf{T}}_i)$.
\item[(3)] Taking the feature vector of the $a$-th frame $\hat{\mathbf{F}}_i$ as input and the acceleration of the corresponding frame in the physical sequence $\mathbf{T}$ as output, construct a new training sample $(\hat{\mathbf{F}}_i,\mathbf{T})$ and include it in the training sample set to adjust and update the weights and biased of the network, completing an iteration.
\item[(4)] Continue the iterative process until the sequence is terminated, resulting in a trained $PcNet$.
\end{enumerate}

\section{3 Experimental results}
The environment for the experiments is as follows: Windows 7 Ultimate 64bits SP1, Intel Xeon E 3-1230 V2@ 3.30-GHz quad-core processor, 8 GB internal storage, and graphics card of NVIDIA GeForces GTX 650 Ti (1 GB).

We choose the GPU accelerated blood vessel wall coupling method proposed by Guo et al.\cite{15} as a physical simulation method for acquiring training samples and as a contrast method for the proposed method. The simulation scenario where we get the input data (the particle state feature vector of the current frame) and the target data (the particle acceleration of the next frame of blood flow) is as follows: Initialize a columnar blood flow of a certain height and let it flow into the fixed blood vessel model under gravity. The time step is 0.005 seconds.
\begin{figure}[H]
\centering
\includegraphics[width=2.5in]{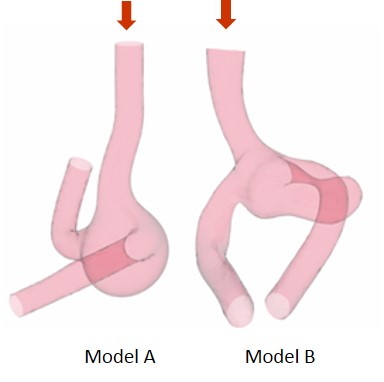}
\caption{Two blood vessel models in experimental setting}
\end{figure}
As shown in Figure 4, we use two blood vessel models to simulate in the above scenario respectively and collect the data to train PcNet with the semi-supervised periodic-corrected training strategy. Taking Model A as an example, five sets of data were obtained from five groups with different initial height simulations. Each set of data collected 800 frames with 16000 particles per frame, for a total of 64 million samples.

Our basic neural network has 24 neurons in the input layer and 3 neurons in the output layer. In the initial testing and verification process, the neural network can describe the nonlinear relationship between input layer and output layer relatively accurately when the number of hidden layers reaches 3 and the number of hidden layer neurons reaches 5. As the number of hidden layers and the number of hidden layer neurons continue to increase, the performance of the network has not significantly improved, consuming a large amount of computing resources at the same time. In this paper, the basic neural network has 3 hidden layers and 5 neurons per hidden layer.

We initialize a new height of columnar blood flow to simulate a test scenario, using the trained PcNet to continuously predict the state of the blood flow. As shown in Figure 5, our method achieves a realistic and stable dynamic simulation of blood flow and vessel wall coupling and achieves almost the same visual effects as the physical method. Figure 6 shows comparison of the coupling simulation results of two methods in a certain frame from different perspectives. Our method even gives more realistic details of blood splashes based on the visual similarity of physics-based method. In addition, the computational efficiency of the proposed method in this paper is about 5 times increase than that of the physics-based complex computations (As shown in Table 1).

\begin{figure*}[tp]
\centering
\includegraphics[width=5in]{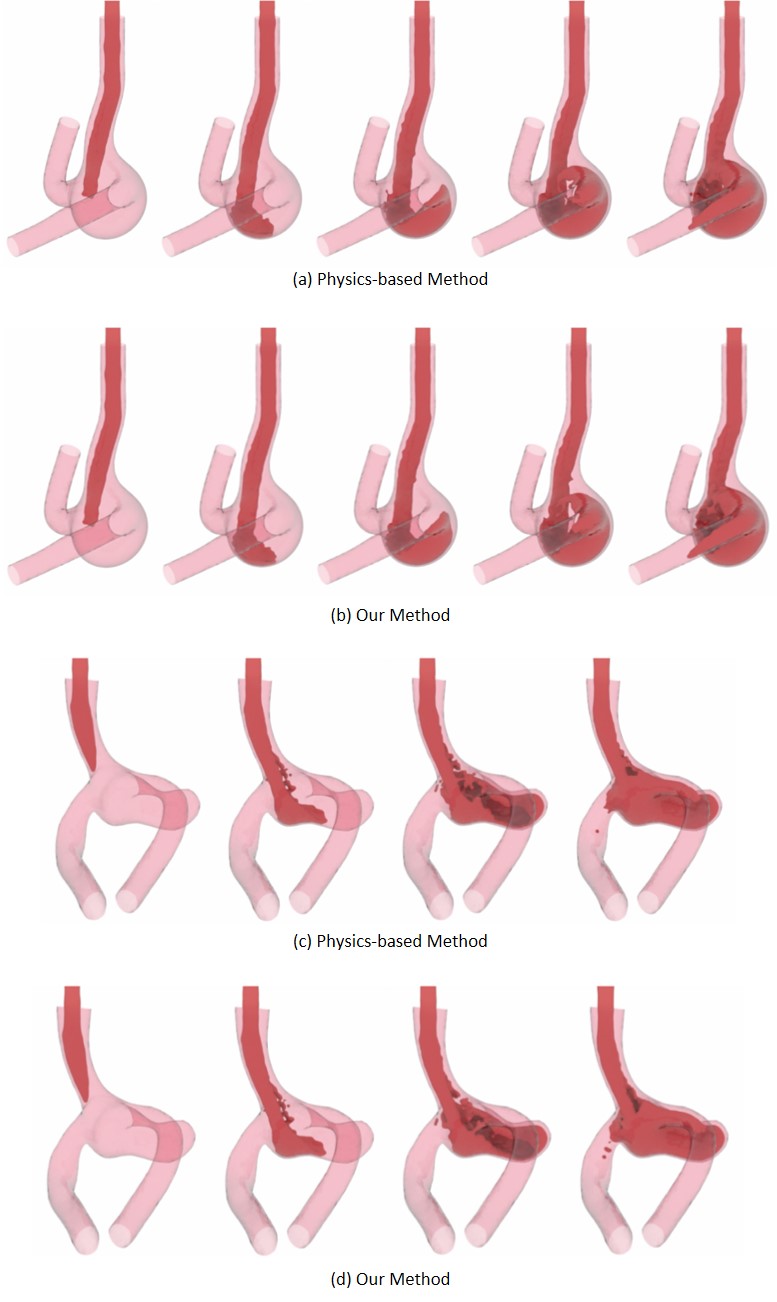}
\caption{Time-series comparison with the simulation results using physical-based method (from left to right, the simulation time is increasing). Our method obtained similar vivid and stable results with some differences.}
\end{figure*}
\begin{table*}[bp]
\small\sf\centering
\caption{Comparison of the physics-based method and our data-driven method in computing time per frame with two models.\label{T1}}
\centering
\begin{tabular}{ccccccc}
\toprule
Model&Blood particles&Proxy particles&All particles&Physical-based method(s)&Our method(s)&Speed-up\\
\midrule
\multirow{2}*{A}& 16000& 56000& 72000& 0.026& 0.0051& 5.1\\
~& 32000& 56000& 88000& 0.033& 0.0063& 5.23\\
\multirow{2}*{B}& 16000& 73000& 89000& 0.03& 0.0064& 4.69\\
~& 32000& 73000& 105000& 0.035& 0.0074& 4.73\\
\bottomrule
\end{tabular}\\[10pt]
\end{table*}
\begin{figure*}[tp]
\centering
\includegraphics[width=5in]{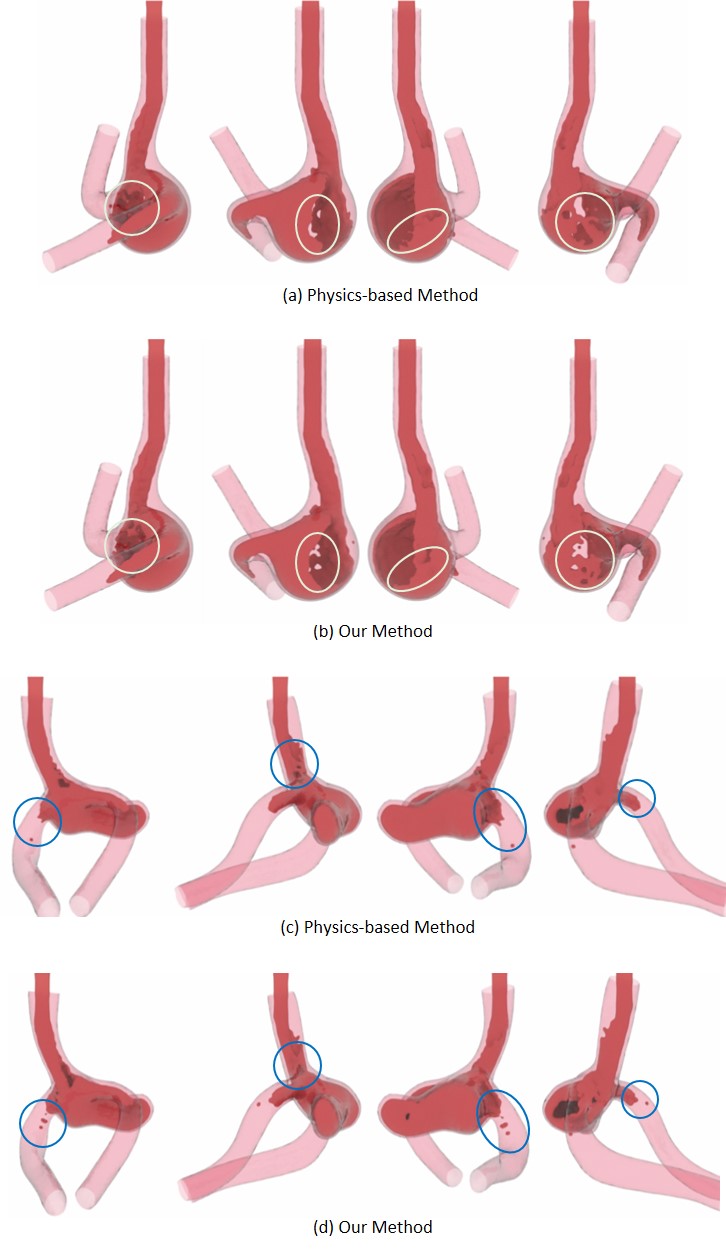}
\caption{Comparison of the simulation results of a particular frame in four different perspectives. Our method obtained more realistic blood splatter details (circled in the figure).}
\end{figure*}
\section{4 Conclusions}
In this paper, we propose an efficient data-driven method to realize fast, stable and visually realistic coupling of blood flow and vessel wall by using periodic-corrected neural network (PcNet) to avoid the complex acceleration computation in physics-based blood flow simulation. Based on the SPH method and the hemodynamic equation, we first construct a particle state feature vector which models the mixed neighboring contribution of proxy particles on the blood vessel wall and blood particles. Then in order to strengthen the physical meaning of the frame sequence to ensure the long-term stability of the simulation, we present a semi-supervised periodic-corrected training strategy to improve the traditional BP neural network. Finally, we train the proposed PcNet and conduct simulation experiments.

The experimental results show that the proposed method achieves almost the same simulation visual effects as the physics-based method, and even better on the details of the blood splashes, while the computational efficiency is improved by about 5 times.
In the future, we will improve feature selection and blood flow incompressibility constraints to support larger scale simulation stability.In addition, we will apply parallel computing such as GPU acceleration to our data-driven method to be more time-saving in training samples and computing coupling results.

\bibliographystyle{SageV}

\end{document}